# Electron diffraction study of the formation and growth of clusters in supersonic binary $N_2$-Kr gas jets

O.P. Konotop, O.G. Danylchenko

B. Verkin Institute for Low Temperature Physics and Engineering of the National Academy of Sciences of Ukraine
47 Nauky Ave., Kharkiv, 61103, Ukraine
e-mail: *konotop@ilt.kharkov.ua*

An electron diffraction diagnostics of substrate-free clusters formed in $N_2$-Kr binary jets expanding through a supersonic nozzle into vacuum was carried out. Gas mixtures contained 0.5, 1, and 6 mol.% krypton, the measured average sizes of aggregations in the cluster beam varied from 500 to 30,000 molecules per cluster. A change in the nucleation mechanism in the jet from heterogeneous to homogeneous was revealed when the temperature of the gas mixture at the nozzle inlet $T_0$ increased from 100 K to 120 K, which had a profound effect on the sizes, phase composition and component composition of the clusters. It was established that the intensification of cluster growth by inserted krypton nucleation centers at $T_0$ = 100 K occurs through an increase in the fraction of the fcc phase. At $T_0$ = 120 K, the effect of cluster growth suppression by the addition of an impurity with significantly stronger intermolecular forces was revealed for the first time. It is shown that the effect is manifested when the krypton gas content increases to 6 mol.% and is caused by the kinetics of gas condensation in a supersonic jet.

## Introduction

Due to simplicity of the forces of interparticle interaction, the clusters of rare and simple molecular gases are common model objects for theoretical modeling and experimental studies of the influence of quantum size effects on physical properties of nanoclusters as well as materials created on their basis. Substrate-free clusters are usually created by adiabatic gas expansion into a vacuum through a supersonic nozzle. In this case, a growth of free clusters is not influenced by the properties of the substrate. When the gas is cooled in the jet, a heat exchange between the gas and the environment is negligible and the gas temperature gets close to its triple point one. These factors facilitate the rapid run of relaxation processes in clusters and the achievement of an equilibrium state of cluster structure. In such systems, it is possible to study cluster effects in the most pristine form, including effects associated with a change in the size of nanoscale aggregations.

When a binary gas mixture is directed to the nozzle inlet, depending on the composition of the mixture, its initial temperature $T_0$, and pressure $P_0$, either homogeneous clusters containing one of the components or heterogeneous aggregations can be formed in the jet. The latter can be considered as nanoscale analogs of bulk solid solutions, alloys, and compounds of solids. In contrast to single-component clusters, their physical properties are determined not only by quantum size effects but also by the concentration of the components and their spatial distribution in the cluster. For example, in the Ar-Xe [1, 2] and Ar-Ne [3] clusters, the radial segregation of the components is observed, i.e. the xenon (argon) core is covered with an argon (neon) shell. Such decay into pure components is not typical of bulk samples [4, 5] and films [6]. Partial radial segregation (when the atoms of the heavier component are located predominantly in the center of the cluster) is also observed in small (~100 molecules) Ar-Kr [7, 8], Kr-Xe [9], and $N_2$- Ar [10] clusters, although these binary systems in the case of large clusters and bulk materials are characterized by good mutual solubility of the components. A distinctive feature of substrate free binary clusters is the so-called enrichment effect, which consists of a significant increase in the concentration of a component with stronger intermolecular forces compared to its content in the initial gas mixture. The formation of the component composition of clusters in binary supersonic jets has received considerable attention in [11, 12, 13]. Some other noteworthy distinctive cluster effects in binary clusters of molecular and rare gases are the intensification of the growth of Ar clusters by the inserted nucleation centers in Ar-Kr gas mixtures [14]; the transformation of the hcp structure in $N_2$-Ar clusters into a mixed hcp + fcc one, caused by an increase in argon concentration, without a sharp change in interatomic distances [15]; the intensification of the hcp phase growth in large (> $10^4$ atoms) binary Ar-Kr clusters compared to single-component argon and krypton clusters [16].

The object of this work is substrate free $N_2$-Kr clusters. Their structure and physical properties have not been studied previously. A characteristic feature of the $N_2$-Kr system is the complete compliance of its components with the Hume-Rothery rules of solid solubility [17], which implies their unlimited mutual solubility. However, according to the phase diagram for bulk alloys [18], both at temperatures below the phase transition in nitrogen (35.6 K) and above it, solid solution decay phases are observed in a wide concentration range. No less complex structural transformations were detected by electron diffraction [19] in $N_2$-Kr thin-film samples condensed onto a substrate at 20 K. Two dual-phase regions were determined in the krypton concentration range of 20–89 mol.%: fcc-Pa3 (20-57 mol.%) and fcc-hcp (58-89 mol.%). It is worth mentioning the work [20] dedicated to the stabilization of atomic nitrogen in impurity-helium condensates (IHC) created by injecting a jet of a gas mixture of $N_2$-Kr-He into superfluid helium. $N_2$-Kr nanoclusters created as part of the IHC under such non-equilibrium conditions presumably have a core-shell structure with krypton atoms forming the core of the cluster. The aim of present work is to define the main features of cluster growth in quasi-equilibrium conditions of binary $N_2$-Kr supersonic jet. In particular, it was necessary to study the influence of the initial concentration of krypton in the gas mixture and the initial temperature $T_0$ of the mixture on the size and structure of clusters.

## Experiments

In this work, we employed the electron diffraction method, which has been widely used over the past decades in studying the structure of free cryocrystal clusters, as well as colloidal particles, surface films, and thin adsorption layers. A detailed description of our experimental technique is given in [21, 22]. The generator of a supersonic cluster beam was hermetically connected to the electron diffraction column. To maintain the working vacuum, a powerful hydrogen condensation pump was located inside the column. The geometric parameters of the supersonic nozzle were as follows: the orifice diameter $d$ was 0.34 mm, the half opening angle of the nozzle $\alpha$ was $4.3^{\circ}$, the exit-to-throat area ratio was 36.7, Mach number at the nozzle exit was 8. Preparation of $N_2$-Kr gas mixtures of a required composition was carried out on an autonomous setup, gas cylinders with components of the mixture were hermetically connected to it. To prepare the mixtures, high-purity gases were used. According to the results of chemical analysis, the total fraction of impurities in pure gases did not exceed $10^{-3}$ %. The concentration composition of the prepared mixtures was controlled by the partial pressure of the components. The average cluster size was changed by varying the pressure $P_0$ in the range from 0.15 MPa to 0.6 MPa while the temperature $T_0$ was fixed. We used two values of $T_0$, one of which (120 K) is above the temperature of the krypton triple point (115.8 K), while the second (100 K) is below it. That choice was conditioned by the expected effects associated with changes in the mechanisms of cluster formation.

Diffraction patterns were recorded by a photographic method with their further scanning and computer processing. This made it possible to obtain a larger set of data in one experiment as well as to reduce significantly the consumption of the studied gases and cryo-liquids used in the vacuum system of the setup. To determine the average linear cluster size $\delta$(Å), the well-known Scherrer relation was used by measuring the broadening of diffraction peaks and taking into account their additional broadening due to stacking faults. The relation between $\delta$ and the average number of particles per cluster $N$ depends on the type of crystal lattice, and for spherical clusters with an fcc structure has the form: $\delta = a(3N/2\pi)^{1/3}$. Here $a$ is the crystal lattice parameter of the cluster, which was determined from the positions of the diffraction peaks. In present measurements, the value of $N$ was varied from $5\cdot10^2$ to $3\cdot10^4$ molecules per cluster. The relative error in the determination of $N$ did not exceed ± 30%.

## Results and discussion

Figure 1 shows sections of electron diffraction patterns for pure $N_2$ clusters and mixed $N_2$-Kr clusters obtained at pressure $P_0$ = 0.5 MPa and initial gas temperatures $T_0$ = 100 K (Fig. 1a) and $T_0$ = 120 K (Fig. 1b). The concentrations of krypton in the initial gas mixtures were 1 mol.%, 6 mol.% for $T_0$ = 120 K, and 0.5 mol.%, 1 mol.% for $T_0$ = 100 K. It should be noted that such low values of the krypton gas concentration at $T_0$ = 100 K were caused by the fact that at higher concentrations krypton condensed on the nozzle walls, which led to a decrease in the effective diameter of the nozzle and even to its complete clogging. As noted before, a one of the peculiarities of the $N_2$-Kr system is the compliance of its components with the Hume-Rothery rules. This compliance is also valid for free clusters, for example, the lattice parameter of the cubic α-phase of nitrogen clusters (5.69 Å)[23] almost coincides with the fcc phase parameter of krypton clusters (5.71 Å)[14]. Due to such closeness of the lattice parameters of krypton and nitrogen, in this study, unfortunately, it was not possible to estimate the concentration of components in $N_2$-Kr clusters by measuring changes in interplanar spacing of the crystalline solid solutions. However, taking into account the data for Ar-Kr and $N_2$-Ar clusters [12], we can confidently assume that the krypton content in $N_2$-Kr clusters was at least 10 times higher than in the initial gas mixtures.

As can be seen from Fig. 1, at krypton gas concentrations of 0.5 mol% and 1 mol% for both fixed temperatures $T_0$, the phase composition of mixed clusters differs slightly from the composition of pure $N_2$ clusters. The position and intensity of diffraction peaks indicate that in both cases the clusters have a dual-phase fcc-hcp structure. The increase in the size of the clusters, either by lowering $T_0$ or by increasing $P_0$, leads to a growth of the relative share of the hexagonal phase. This phenomenon is caused by increasing the time the cluster is in a heated state at premelting temperatures, which boosts the rate of transition of the cluster to equilibrium state. At the same time, it was found that the addition of krypton to nitrogen leads to a decrease in the fraction of hcp phase compared to the fcc one. Such a structural transformation has previously been observed in free $N_2$-Ar clusters when argon is added to nitrogen [4], but it is opposite to the effect of hcp phase growth intensification in mixed Ar-Kr clusters [16]. As can be seen from Fig. 1b, the smallest fraction of hexagonal phase in the clusters was observed at $T_0$ = 120 K and an increase in the krypton gas content to 6 mol.%. It was found that in this case at low values of $P_0$ the hcp phase in the clusters is absent, and with an increase in pressure (and hence in the mean cluster size), the structure is transformed in the following way: icosahedral → fcc with stacking faults → fcc + hcp. Such transformation is inherent in pure rare gas clusters [3], that gives us a reason to assume that at a krypton content in the $N_2$-Kr gas mixture of 6 mol% the cluster beam contains only krypton aggregations. It is discussed below whether or not this is the case.

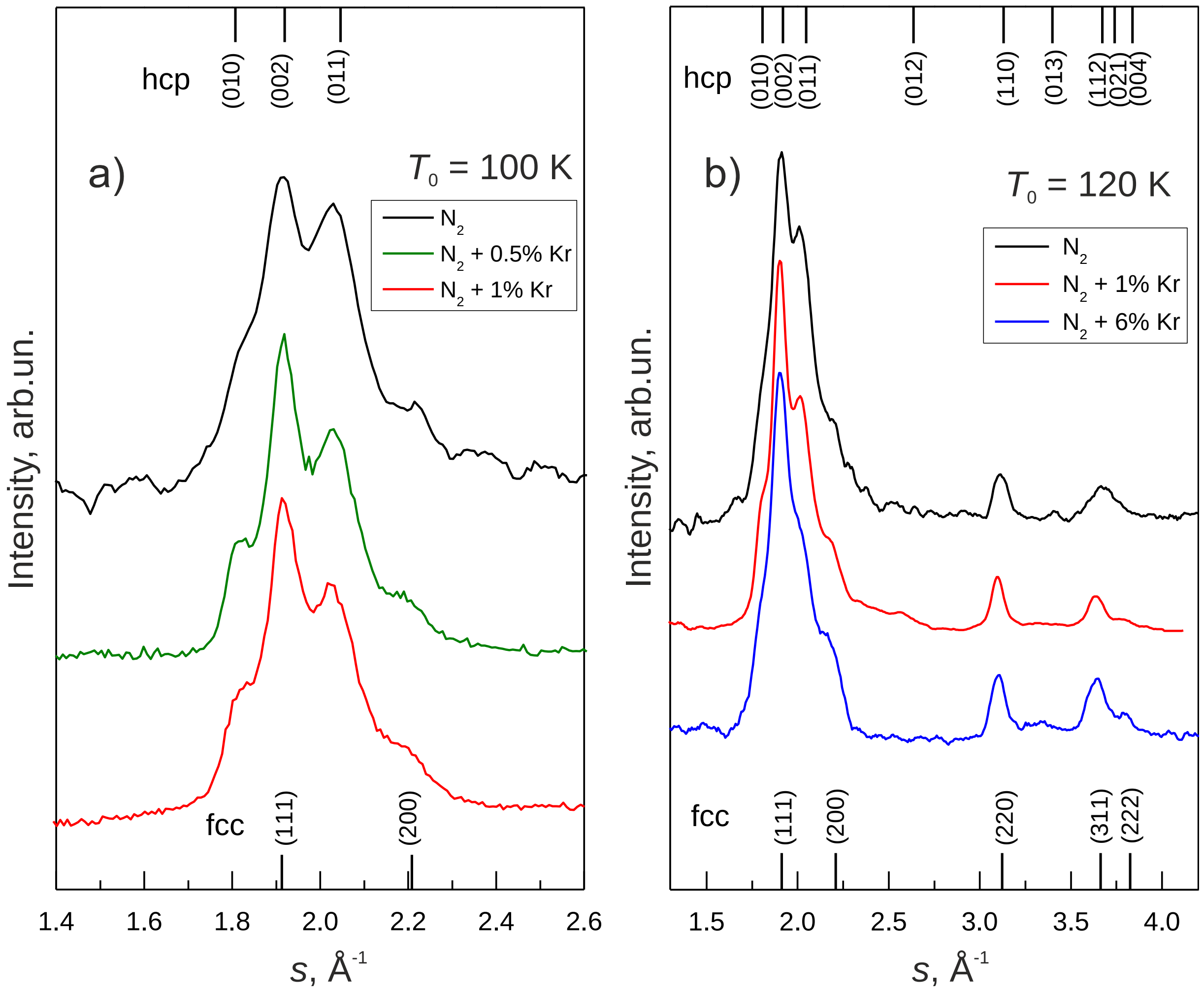

Fig.1. Electron diffraction patterns of pure $N_2$ clusters and clusters formed in binary $N_2$-Kr gas jets of various krypton concentrations. Initial gas pressure $P_0$ is 0.5 MPa, initial gas temperature $T_0$ is 100 K (a) and 120 K (b).

Fig. 2 presents the dependences of the average cluster size $N$ on pressure $P_0$ for pure $N_2$ clusters and mixed $N_2$-Kr clusters. The approximating curves correspond to the theoretical dependences of $N \sim P_0^{1.8}$, which are valid for the nozzle used in present study [24, 25]. It is clear from the experimental data that the addition of krypton to nitrogen at different temperatures $T_0$ can lead to either the intensification of cluster growth or its suppression. And while the former effect was previously observed in Ar-Kr clusters [14], the latter is rather unexpected and has not been observed before. Let's consider these effects one by one.

At a temperature $T_0$ = 100 K (Fig. 2a), the addition of a small portion of krypton to the initial nitrogen gas results in a twofold increase in the average number of molecules per cluster. It should be noted that in Ar-Kr clusters under identical experimental conditions, the intensification of cluster growth is even more significant and is characterized by an increase in *N* by almost an order of magnitude. In both cases, the mechanism of cluster growth intensification is the realization of heterogeneous (barrier-free) clustering of primary gas molecules on krypton nanoparticles, since condensation of krypton at such $T_0$ occurs even before reaching the nozzle critical cross-section. Thus, the presence of impurities acts in the same way as increasing the pressure $P_0$ at the nozzle entrance, i.e. it increases the size of the aggregations. However, these increases have different mechanisms, which are reflected in the structure of $N_2$-Kr clusters. As it was mentioned above, the growth of clusters due to the change in $P_0$ is accompanied by an increase in the share of the hcp phase. On the contrary, the intensification of cluster growth on the inserted nucleus centers leads to an increase in the share of the fcc phase (see Fig. 1a). The reason for this phenomenon probably lies in the crystal structure of krypton nanoparticles, which form the core of mixed clusters. Small krypton clusters are characterized by an fcc or fcc-like icosahedral structure, therefore, when nitrogen is heterogeneously clustered on krypton nuclei, it is easier for $N_2$ molecules to occupy the positions of a cubic lattice rather than a hexagonal one.

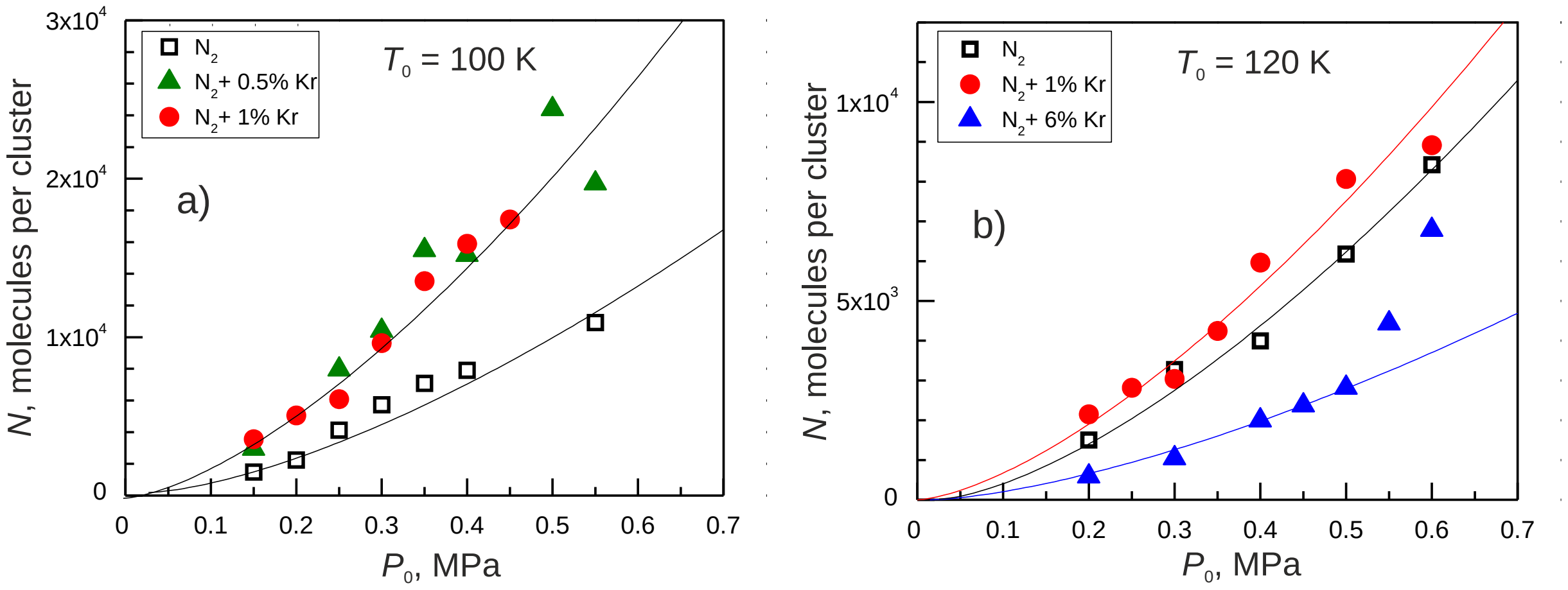

Fig.2. Dependences of the average cluster size *N* on initial gas pressure $P_0$ for pure $N_2$ clusters and clusters formed in binary $N_2$-Kr gas jets of various krypton concentrations. Initial gas temperature $T_0$ is 100 K (a) and 120 K (b).

As the temperature $T_0$ increases to 120 K, the probability of heterogeneous clustering decreases sharply, as shown in Fig. 2b. At a krypton gas content of 1 mol.%, the addition of the impurity to nitrogen leads to only a slight increase in the size of the clusters, which is smaller than the measurement error of *N*. Moreover, when the krypton gas content increases to 6 mol.%, a significant decrease in the average size of the clusters with respect to pure nitrogen clusters (i.e. cluster growth suppression) is observed. Only at $P_0 > 0.5$ MPa does the $N(P_0)$ dependence deviate upwards from the theoretical $N \sim P_0^{1.8}$ and approaches the $N(P_0)$ dependence for pure nitrogen clusters. Suppression of cluster growth by adding impurity with significantly stronger intermolecular forces and triple point temperature to the jet is a highly unusual effect that has not been observed experimentally before. One of the probable reasons for this effect that we considered was a technical limitation, namely a decrease in nozzle performance due to the condensation of krypton on its surface. However, in this case, an increase in $P_0$ would have not led to an experimentally measured increase in cluster sizes, which corresponds with the theoretical dependence $N \sim P_0^{1.8}$, but on the contrary, it would have led to a complete clogging of the nozzle.

To identify the source of cardinal difference in $N(P_0)$ dependences for $N_2$-Kr beams with different krypton gas content at $T_0$ = 120 K, the average distances $z$ between the centers of the nearest molecules in the crystal lattice of the studied clusters were determined (Fig. 3). An increase in $z$ with rising pressure is observed both in pure $N_2$ clusters and in mixed ones. It is associated with an increase in the hcp phase share in the clusters, as it is known that the transition from the low-temperature cubic α-phase to the high-temperature hexagonal β-phase in bulk nitrogen cryocrystals is accompanied by a slight enlargement of intermolecular distances [4]. It can also be seen from Fig. 3 that the value of $z$ for 1 mol.% Kr in the gas mixture is slightly greater than its value for pure nitrogen clusters. This result indicates the formation of substitutional solid solutions. On the contrary, a decrease in $z$ for clusters of the $N_2$ + 6 mol.% Kr mixture compared to $z$ for pure $N_2$ clusters is observed. This $z$ decrease is an additional argument for our suggestion about formation of only one-component krypton clusters in the binary jet, despite the fact that the value of $z$ in clusters formed in a supersonic jet of pure krypton (dashed line in Fig. 3) is significantly larger. The point is that at krypton gas content of 6 mol.%, nitrogen molecules act as a buffer gas, thus contributing to faster cooling of condensed krypton aggregations. Due to this, the temperature of krypton clusters formed in the $N_2$-Kr mixture ($T_{cl} \approx 40$ K, according to the temperature dependence of the lattice parameter for bulk Kr [4]) is significantly lower than the temperature of clusters formed in pure krypton jets ($T_{cl} \approx 60$ K [14]). In turn, a decrease in the temperature of the clusters is accompanied by a decrease in the crystal lattice parameters. As a result, the intermolecular distances z in the Kr clusters formed in the binary $N_2$ + 6 mol.% Kr jet are significantly smaller than those in the clusters of pure krypton jet.

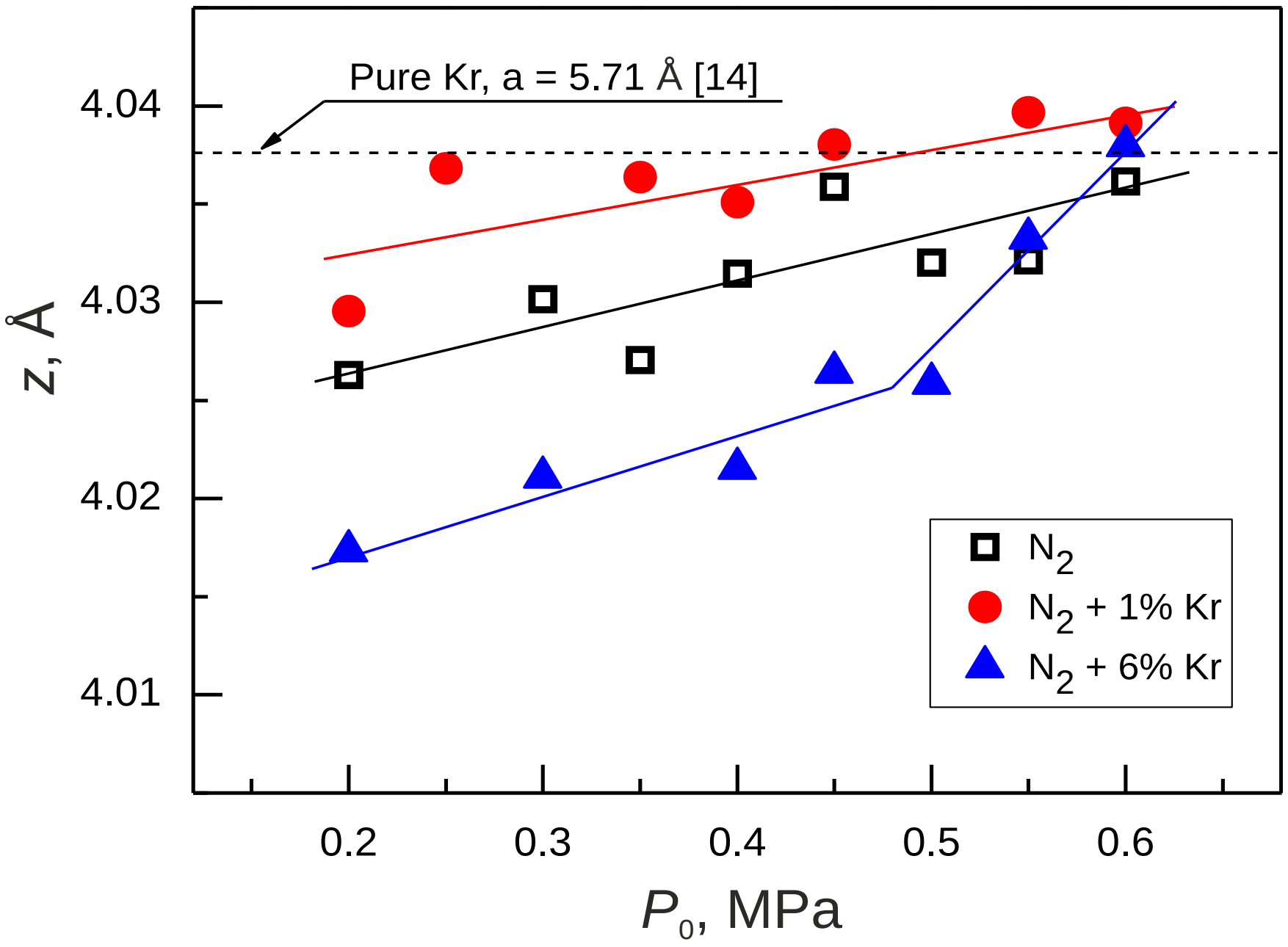

Fig.3. Dependences of the average distance $z$ between the centers of the nearest molecules in the crystal lattice on initial gas pressure $P_0$ for pure $N_2$ clusters and clusters formed in $N_2$ + 1 mol.% Kr and $N_2$ + 6 mol.% Kr gas mixtures. Initial gas temperature $T_0$ is 120 K. The dashed line indicates the $z$ value for pure krypton clusters [14].

Therefore, the most likely reason for the effect of cluster growth suppression in $N_2$ + 6 mol.% Kr mixture is the impossibility of condensation of nitrogen molecules at a relatively high concentration of krypton atoms in the jet. An obstacle to nitrogen condensation is the peculiarity of the kinetics of cluster formation in a supersonic jet, namely the release of the latent heat of condensation of krypton. The specific heat of condensation of krypton (9.03 kJ*mol$^{-1}$) is significantly higher than the one of nitrogen (2.79 kJ*mol$^{-1}$). Its release leads to a sharp increase in the gas jet temperature, and if the krypton content in the mixture is high enough, the jet temperature becomes too high for nitrogen condensation. Only by increasing the initial pressure $P_0$, it is possible to create such supersaturation conditions in the jet, that become sufficient for the condensation of nitrogen molecules, this is evidenced by a sharp increase in the intermolecular distances $z$ (Fig. 3) and the average sizes of clusters $N$ (Fig. 2b) at $P_0 > 0.5$ MPa .

## Conclusions

For the first time, an electron diffraction study of the formation and growth of clusters in supersonic binary $N_2$-Kr gas jets expanding into a vacuum was carried out. The krypton content in gas mixtures was equal to 0.5, 1 and 6 mol.%. The studies were performed with a change in the total pressure of the mixture at the nozzle entrance $P_0$ from 0.15 to 0.6 MPa, and two initial gas temperatures $T_0$: 100 K (i.e. below the krypton triple point) and 120 K (above it). Depending on the experimental conditions, the measured average size $N$ of the formed aggregations varied from 500 to 30,000 molecules per cluster.

In crystalline clusters (N > 2000), the coexistence of fcc and hcp structures was observed, while the relative fraction of the fcc phase increased with the increase in the krypton concentration in the mixture. An enlargement of the cluster sizes by a change of the initial thermodynamic parameters of the mixture, on the contrary, led to an increase in the relative fraction of the hexagonal phase.

It was established that the mechanisms of cluster formation can be fundamentally different depending on the temperature of the gas mixture at the nozzle entrance and the concentration of krypton impurity. In particular, the addition of krypton to the nitrogen jet can lead to either the intensification of cluster growth or its suppression:

- When $T_0$ is below the krypton triple point and the krypton content in the mixture is up to 1 mol.%, cluster growth intensifies because of the insertion of preliminary condensed krypton nanoparticles into the supersonic region of the nozzle. The mechanism of cluster formation is heterogeneous and similar to that implemented in a pick-up technique of binary clusters creation. As a result, the average size of the clusters doubles compared to pure nitrogen aggregations. This size doubling, in contrast to the growth of clusters caused by the change of $T_0$ and $P_0$, is accompanied by the growth of the fcc phase fraction, not the hcp one.

- When $T_0$ is above the krypton triple point and the krypton content in the mixture is up to 1 mol.%, condensation of krypton molecules occurs only after they pass through the nozzle throat. Relatively low concentration of Kr in the jet does not prevent nitrogen condensation and leads to the formation of mixed $N_2$-Kr clusters, which have the structure of substitutional solid solutions and are quite similar in size and phase composition to pure $N_2$ clusters.

-When $T_0$ is above the krypton triple point and the krypton concentration in the mixture is more than 5 mol%, the effect of cluster growth suppression by a large content of impurity with significantly stronger intermolecular forces was detected for the first time. Due to the huge release of the latent heat of condensation of krypton, its fraction in the mixture of 6 mol.% turned out to be too large for the condensation of nitrogen molecules to occur, but at the same time it is too small for the krypton clusters formed in the binary jet to catch up in size with the clusters of pure nitrogen formed at the same.